\documentclass[11pt,a4paper]{amsart}

\usepackage[latin1]{inputenc}
\usepackage[english]{babel}
\usepackage{amsmath}

\usepackage{amssymb, mathabx}
\usepackage[numbers]{natbib}
\usepackage{graphicx}
\usepackage{braket}
\usepackage[colorlinks,hyperindex,bookmarksopen,linkcolor=red,citecolor=blue,urlcolor=blue]{hyperref}

\usepackage{mathrsfs}

\usepackage{epstopdf}

\usepackage{braket}

\bibpunct{[}{]}{;}{n}{,}{,}
\usepackage[hmargin=3cm,vmargin={3.5cm,4cm}]{geometry}

\theoremstyle{theorem}

\theoremstyle{definition}                                 

\theoremstyle{definition}                           

\theoremstyle{remark}                             

\usepackage{color}

\usepackage{mathtools,slashed}

\newcommand{\be}{\begin{eqnarray}}
\newcommand{\ee}{\end{eqnarray}}

\numberwithin{equation}{section}

\allowdisplaybreaks

\begin{document}

\title{Invasion moving boundary problem for a biofilm reactor model}

\author{B. D'Acunto}

\author{L. Frunzo}

\author{V. Luongo}

\author{M. R. Mattei}

\address{University of Naples "Federico II", Department of Mathematics and
					 Applications, Complesso Monte Sant'Angelo, 80124, Naples, Italy\
          \texttt{dacunto@unina.it, luigi.frunzo@unina.it, vincenzo.luongo@unina.it, mariarosaria.mattei@unina.it} }


\date  {\today}

\begin{abstract}

The work presents the analysis of the free boundary
 value problem related to the invasion model of new species in biofilm reactors.
 In the framework of continuum approach to mathematical modelling
 of biofilm growth, the problem consists of a system of nonlinear
 hyperbolic partial differential equations governing the microbial
 species growth and a system of semi-linear elliptic partial
 differential equations describing the substrate trends. The model
 is completed with a system of elliptic partial differential
 equations governing the diffusion and reaction of planktonic
 cells, which are able to switch their mode of growth from
 planktonic to sessile when specific environmental conditions are
 found. Two systems of nonlinear differential equations for the
 substrate and planktonic cells mass balance within the bulk liquid
 are also considered. The free boundary evolution is governed by a
 differential equation that accounts for detachment.
 The qualitative analysis is performed and a uniqueness and 
 existence result is discussed.
 Furthermore, two special models of biological and engineering
 interest are discussed numerically. The invasion of Anammox bacteria in a constituted
 biofilm inhabiting the deammonification units of the wastewater treatment
 plants is simulated. Numerical simulations are run to evaluate 
 the influence of the colonization process 
 on biofilm structure and activity.

\end{abstract}
\maketitle

\section{Introduction} \label{n1}
 \setcounter{equation}{0} \setcounter{figure}{0}

 The term biofilm is used nowadays to indicate the prevailing form
 of microbial lifestyle, which consists of dynamic complex
 microbial structures composed of various prokaryotic cells and
 other microorganisms, forming on solid or liquid surfaces and
 encased in a self-produced protective matrix of extracellular
 polymeric substances (EPS). The roles biofilms exert on both
 natural and human environments are disparate: they have proven
 detrimental to human health or undesirable in the open water
 environment but, on the other hand they can be used beneficially
 in resource recovery systems as well as water treatment
 \cite{boltz2017}. With specific reference to the last point,
 biofilm reactors represent the primary means to harness the
 usefulness of biofilms for pollutant removal from wastewater by
 means of the synergistic interactions and biochemical
 transformations characterizing these microbial communities
 \cite{mattei2}. The biofilm structure results from the interplay
 of different interactions, such as mass transfer, conversion rates
 and detachment forces. The main biofilm expansion is due to
 bacterial growth and to extracellular polymer production. The soluble substrates necessary for bacterial
 growth are dissolved in the liquid flow and to reach the cells,
 first they pass through a boundary layer, characterized by a
 negligible flow over the biofilm/liquid interface, and then
 through the biofilm matrix. The external fluid flow regulates
 biofilm growth by establishing the concentration of substrates and
 products at the solid-liquid interface and shearing the biofilm
 surface. Other biological phenomena are found to play
 significant roles in the establishment of mixed species biofilms,
 i.e. dispersal, bacteriophage, quorum sensing \cite{ward}.

 Among these phenomena, there is a growing interest in the study of microbial invasion and
 colonization of pre-existing biofilms as
 it can determine biofilm landscape and contribute to rapid alterations in biofilm populations. 
 Recent advances in microbial ecology have identified
 motility as one of the main mediator of such process. Indeed, once a motile bacteria,
 supplied by the liquid phase or the biofilm itself (as a consequence of dispersal phenomenon),
 has successfully infiltrated the biofilm matrix, it can invade a resident community and establish where the environmental
 conditions are optimal for its growth. An accurate modeling of such a system has to take all of these factors into account.
 In a recent contribution \cite{inv}, the authors have introduced a multispecies biofilm model which explicitly takes into account the invasion phenomenon
 pursued by planktonic cells. The core of the model lies on the introduction of new state variables which represent the concentrations of planktonic colonizing cells
 within the biofilm. These cells are supposed to be characterized by a diffusive movement within the biofilm and to be able to give up
 the ability to move in order to settle down in specific environmental niches.

 In this work, we introduce the free boundary value problem for the invasion phenomenon in biofilm reactors which takes into account the dynamics
 of the bulk liquid phase in terms of both substrates and planktonic cells. The mathematical problem consists of a system of
 hyperbolic partial differential equations governing the biofilm growth, a system of elliptic partial differential equations
 for substrate dynamics within the biofilm and a system of elliptic partial differential equations, regulating the diffusion and
 reaction of planktonic cells. Mass balance equations for the dissolved substrates and planktonic cells within the bulk liquid phase
 of the biofilm reactor have been taken into account as well. The free boundary evolution is governed by a nonlinear ordinary differential equation.

 The qualitative analysis of such a complex system is not an easy task as outlined in \cite{masic}. Due to the high non-linearity of the problem,
 the fixed point theorem seems the natural tool to be used for the existence and uniqueness of the solutions. However, we are considering a moving
 boundary problem where the domain is not fixed. To overcome this issue, we follow the methodology used
 in \cite{reactor} for the analysis of the biofilm reactor model and in \cite{inv,inv2} for the modeling
 of the planktonic cells dynamics both within the biofilm and the bulk phase. In particular, we use the method of characteristics to convert
 the differential problem to an integral one where the unknown functions are defined on a fixed
 domain and the existence and uniqueness of the solutions are proved in the class of continuous functions.

 In addition, the work is completed with some numerical applications related to a real
 engineering/biological case which examines the invasion of specific microbial species in a constituted biofilm. 
 More precisely, the case study reproduces the invasion of Anammox bacteria within a multispecies biofilm 
 devoted to the concurrent oxidization of ammonium nitrogen and organic carbon occurring in the biological units
 of the wastewater treatment plants. Traditionally, ammonium oxidation leads to the formation of residual nitrogen compounds that need to be further removed by means of 
 other treatment phases. The establishment of a biofilm community constituted by Anammox bacteria and Aerobic ammonium oxidizers may lead instead to the complete conversion
 of ammonium nitrogen to nitrogen gas within a single treatment unit. The establishment of this syntrophy is catalyzed by the formation of an anoxic zone, where the Anammox
 bacteria can effectively proliferate.
 The invasion model has been adopted to illustrate the trends related to the establishment of such a multispecies community and to assess the effect of specific
 operational conditions on the biofilm colonization by Anammox bacteria.
 For all the cases analyzed real data from existing literature are used to feed
 numerical simulations, which produce results in nice agreement with experimental findings.

 The paper is organized as follows.
 In Section \ref{n2} the invasion moving boundary problem for a biofilm reactor model is introduced: assumptions, boundary and initial conditions are discussed. 
 Section \ref{n3} introduces the Volterra integral equations. Section \ref{n4} describes the experimental case to which the model is applied
 and presents the numerical results. Finally, in Section \ref{n6} we present the conclusions and the future recommendations of the work. 


 \section{Invasion boundary problem for biofilm reactors}   \label{n2}
 \setcounter{equation}{0} \setcounter{figure}{0}

 We analyze the free boundary value problem related to the invasion problem in biofilm
 reactors. In this model we consider the biofilm as constituted by various particulate
 components (i.e. bacteria, EPS, etc.) growing in a liquid environment, and planktonic cells
 belonging to various microbial species and able to move within the biofilm and the bulk liquid as well. 
 The biofilm expansion depends on growth limiting nutrients which are dissolved in the liquid region
 or produced within the biofilm itself. The planktonic cells can
 diffuse from the bulk liquid to the biofilm, invade, and switch their mode of
 growth from suspended to sessile when appropriate environmental conditions are found.

 The model is formulated for the variables concentration of microbial species in
 sessile form $X_i$, the concentration of planktonic invading cells $\Psi_i$, the concentration of
 dissolved substrates $S_j$, all expressed as functions of time $t$ and $z$
 which denotes the spatial coordinate. The substratum is assumed to be placed at $z=0$.
 The one-dimensional form of the model writes

   \begin{equation}                                        \label{2.1}
     \frac{\partial X_i}{\partial t} +\frac{\partial}{\partial z}(u X_i)
     =\rho_i r_{M,i}^{}(z,t,\textbf{X},\textbf{S})+\rho_i r_i^{}(z,t,\textbf{S},
    \mbox{\boldmath $\Psi$}), \ i=1,...,n,
  \end{equation}
	
 where $\rho_i$ denotes the constant density of species $i$, $u(z,t)$
 the biomass velocity  at which the microbial mass is
 displaced with respect to the film-support,
 and $\bf{X}=(X_1,...,X_n)$, $\bf{S}=(S_1,...,S_m)$,
 $\mbox{\boldmath $\Psi$}=(\Psi_1,...,\Psi_n)$.
 The reaction terms
 $r_{M,i}$ describe the
 growth of sessile cells, which is controlled by the local
 availability of nutrients and usually modelled as standard Monod
 kinetics, and natural cell death. In most biological processes
 the function $r_{M,i}$ depends on $z,t$ only trough the functions
 $\bf{X}$ and $\bf{S}$. The explicit dependence has been
 considered mainly for mathematical generality. The variable $t$ is positive
 and $0\leq z\leq L(t)$, where $L(t)$ denotes the biofilm thickness at time $t$.
 Equation (\ref{2.1}) without the term $r_{i}$ was first derived in
 \cite{wanner86} by mass balance principle. The initial conditions for
 (\ref{2.1}) are provided by the initial concentrations
 $\varphi_{i}(z)$ of biofilm particulate components
 \begin{equation}                                        \label{2.2}
 X_i(z,0)=\varphi_{i}(z),\   i=1,...,n, \  0\leq z\leq L(0).
 \end{equation}
 The initial concentrations of the invading microbial species
 are set to zero. The equation in the form (\ref{2.1}) was presented in \cite{inv}.
 The terms $r_i$ represent the growth rates of the microbial species $X_i$ due to the invasion process
 which induces the switch of planktonic cells to a sessile mode
 of growth. This phenotypic alteration is catalyzed by the formation within
 the biofilm matrix of specific environmental niches. The
 explicit dependence on $z,t$ has been introduced only
 for mathematical generality.

 Similarly to traditional continuum models of biofilm growth, equations
 (\ref{2.1}) can be rewritten in terms of volume fractions $f_i=X_i/\rho_i$,
 which indicate the fraction of space at a particular location that is occupied by species $i$,

 \[
     \frac{\partial f_i}{\partial t} +\frac{\partial}{\partial z}(u f_i)
     = r_{M,i}^{}+ r_i^{}.
 \]

 Of course, the sum of all volume fractions at each location and time must always sum to one $\sum_{i=1}^n f_i=1$.

From the equations above it follows immediately that the function
$u(z,t)$ satisfies the following problem

\begin{equation}                                        \label{2.3}
     \frac{\partial u}{\partial z} = \sum_{i=1}^{n}
     \left(r_{M,i}(z,t,{\bf X},{\bf S}) +r_i(z,t,{\bf S},\mbox{\boldmath $\Psi$})\right),
     \ 0 < z\leq L(t), \
     u(0,t)=0,
\end{equation}

\noindent
where the initial condition $u(0,t)=0$ comes from no flux condition on substratum.

The function $L(t)$ is solution of the following problem
\begin{equation}                                        \label{2.4}
      \dot L(t)=u(L(t),t)-\sigma_d(L(t)),\
        L(0)=L_0.
\end{equation}
 Therefore, it is apparent that
        the evolution of the free boundary depends
        on the displacement
        velocity of microbial biomass $u$ and
        detachment flux  $\sigma_d$ as well.
        Equation in
        (\ref{2.4}) comes from global mass conservation principle.
				
 The diffusion of planktonic cells within the biofilm matrix is governed by the
  following diffusion-reaction equations
	
  \begin{equation}                                        \label{2.5}
      \frac{\partial \Psi_i}{\partial t}-\frac{\partial}{\partial z}\left(
        D_{M,i}\frac{\partial \Psi_i}{\partial z}\right)=
        r_{\Psi,i}^{}(z,t,{\bf S},\mbox{\boldmath $\Psi$}), \ i=1,...,n,
        \ 0< z< L(t),
 \end{equation}

 \noindent
 where the reaction terms $r_{\Psi,i}^{}$  represent loss
 terms for the invading species.
 Homogeneous Neumann conditions are adopted on the substratum at $z=0$ due to
 a no-flux conditions
 and Dirichlet boundary conditions are prescribed on the free boundary $z=L(t)$

 \begin{equation}                                        \label{2.6}
  \frac{\partial \Psi_i}{\partial z}(0,t)=0,\
        \Psi_i(L(t),t)=\psi_i^*(t), \ i=1,...,n.
 \end{equation}

 The initial conditions are set to zero if it is assumed that the invasion process
 starts at $t=0$, but specific functions can also be considered.

 The functions $\psi_i^*(t)$ denote the concentrations of planktonic cells
 within the bulk liquid and are governed by the following initial
 value problem for ordinary differential equations

 \begin{equation}                                        \label{2.7}
     V\dot \psi_i^*=-AD_{M,i}^{}
   \frac{\partial \Psi_i}{\partial z}(L(t),t)+
   Q(\psi_i^{in}-\psi_i^*(t)),\ \psi_i^*(0)=\psi_i^{in}, \
    i=1,...,n.
 \end{equation}

 Equations (\ref{2.7}) come from a mass balance within the bulk liquid and account
 for the inlet and outlet flux
 to the biofilm reactor and the exchange fluxes to  or
 from  the biofilm as well. The bulk liquid is modelled
 as a completely mixed compartment of volume $V$ and continuously fed and
 withdrawn at the same flow rate $Q$. The initial concentrations of planktonic
 cells within the bulk liquid
 are provided by the inlet concentrations $\psi_i^{in}$.

 The substrate diffusion within the biofilm is governed by the
 following reaction-diffu\-sion equations

 \begin{equation}                                        \label{2.8}
      \frac{\partial S_j}{\partial t}-\frac{\partial}{\partial z}\left(
        D_{j}\frac{\partial S_j}{\partial z}\right)=
        r_{S,j}(z,t,{\bf X},{\bf S}), \ j=1,...,m, \ 0< z< L(t),
 \end{equation}

 \noindent
 where the terms $r_{S,j}$ represent the substrate production or
        consumption rates due to microbial metabolism and $D_{j}$ denotes the
        diffusion coefficient of substrate j within the biofilm.
 As to the boundary conditions it is assumed that

 \begin{equation}                                        \label{2.9}
    \frac{\partial S_j}{\partial z}(0,t)=0,\
     h\frac{D_{j}}{D_{j}^*}\frac{\partial S_j}{\partial z}(L(t),t) +
 S_j(L(t),t)=S_j^*(t),\ j=1,...,m.
 \end{equation}

 The first conditions is a no-flux boundary condition on the substratum
 placed at $z=0$. The second condition derives from the following
 reasonings. According to \cite{kla} we assume that at a certain distance from
 the substratum $H(t)=L(t)+h$, with $h$ being a given positive
 constant, the substrate concentration $S_j(H(t),t)$ is the same as the bulk
 liquid concentration denoted by $S_j^*(t)$.
  This dissolved substrate diffuses from the bulk liquid to the biofilm
   $0\leq z \leq L(t)$
    where it is consumed   according to equations (\ref{2.8}). No biochemical
    reactions are supposed to occur
    for $L(t) \leq z \leq H(t)$ which leads to consider homogeneous parabolic
        equations for $S_j(z,t)$. Solving at steady-state leads to  (\ref{2.9}),
  where  $D_{j}^*$
        represents the diffusion coefficient of substrate $j$ within the bulk liquid.
  Note that,
        condition (\ref{2.9}) reduces to $S_j(L(t),t)=S_j^*(t)$ for $h=0$.
				
	The functions $S_j^*(t)$ are governed by the following initial
 value problem for ordinary differential equations

 \begin{equation}                                        \label{2.10}
     V\dot S_j^*=-AD_{j} \frac{\partial S_j}{\partial z}(L(t),t)+
   Q(S_j^{in}-S_j^*(t)),\    j=1,...,m,\    S_j^*(0)=S_j^{in}.
 \end{equation}

 Equations above are derived from mass balance on
 the bulk liquid taking into account the inlet and outlet
 flux from the reactor and the exchange flux between
 the biofilm and the bulk liquid. The initial conditions
 for $S_j^*$ are the same as the inlet concentrations.

 Finally, due to the slow evolution of the system \cite{reactor},
 $S_j(z,t)$  profiles can be considered to evolve quasi-statically
 and thus equations (\ref{2.8}) are rewritten as

 \begin{equation}                                        \label{2.11}
        -D_{j}\frac{\partial^2 S_j}{\partial z^2}=
        r_{S,j}(z,{\bf X},{\bf S}), \ j=1,...,m, \ 0< z<L(t),
 \end{equation}

 with boundary conditions  (\ref{2.9}). In addition, same
 arguments as before lead to replace  equations (\ref{2.5}) with
 the following

 \begin{equation}                                        \label{2.12}
        -D_{M,i}^{}\frac{\partial^2 \Psi_i}{\partial z^2} =
        r_{\Psi,i}(z,{\bf S},\mbox{\boldmath $\Psi$}),
        \ i=1,...,n, \ 0< z <L(t),
 \end{equation}

 with boundary conditions  (\ref{2.6}).

 In conclusion the invasion free boundary problem for biofilm reactor is expressed by
 equations (\ref{2.1})-(\ref{2.12}). In the next section,
 following \cite{reactor,inv2}, an equivalent integral formulation of the
 problem will be provided. As it will be apparent at the end of
 the following section, the integral form of the free boundary
 problem presents the great advantage that the space variable is
 defined on a fixed domain whereas in the differential formulation
 (\ref{2.1})-(\ref{2.12}) the space variable belongs to the
 moving domain $0\leq z\leq L(t)$.
	

 \section{Volterra integral equations} \label{n3}
\setcounter{equation}{0} \setcounter{figure}{0}

The differential problem introduced in the previous section is
 herein converted to Volterra integral equations by using the
 method of characteristics. The character\-istic-like lines of
 system (\ref{2.1}) are defined by

 \begin{equation}                                        \label{3.1}
   \frac{\partial c}{\partial t}(z_0^{},t)=u(c(z_0^{},t),t),\
   c(z_0^{},0)=z_0^{},\ 0 \leq z_0^{} \leq L_0,\ t>0,
 \end{equation}
Considering (\ref{3.1}), equations (\ref{2.1}) are converted to
\[
      \frac{d }{d t}X_i(c(z_0^{},t),t)=
 \]
 \begin{equation}                                        \label{3.2}
      F_i(c(z_0^{},t),t,{\bf X}(c(z_0^{},t),t),{\bf S}(c(z_0^{},t),t),
      \mbox{\boldmath $\Psi$}(c(z_0^{},t),t)),
 \  0 \leq z_0 \leq L_0,\ t>0,
 \end{equation}
 with
 \[
    F_i=\rho_ir_{M,i}^{}(c(z_0^{},t),t,{\bf X}(c(z_0^{},t),t),{\bf S}(c(z_0^{},t),t))
 \]
 \begin{equation}                                        \label{3.3}
 +
    \rho_ir_{i}^{}(c(z_0^{},t),t,{\bf S}(c(z_0^{},t),t),
    \mbox{\boldmath $\Psi$}(c(z_0^{},t),t))
   - X_i(c(z_0^{},t),t)\sum_{i=1}^{n}\left(r_{M,i}+r_{i}\right),
 \end{equation}
    and initial conditions
 \begin{equation}                                        \label{3.4}
      X_i(c(z_0^{},0),0)=\varphi_{i}(z_0^{}),
 \ \  0 \leq z_0 \leq L_0.
 \end{equation}
Integrating (\ref{3.2}) and considering (\ref{3.4}) yields
 \[
      X_i(c(z_0^{},t),t)\hspace{-.5mm}=\hspace{-1mm}
 \int_0^tF_i(c(z_0^{},\tau),\tau,{\bf X}(c(z_0^{},\tau),\tau),
 {\bf S}(c(z_0^{},\tau),\tau),\mbox{\boldmath $\Psi$}(c(z_0^{},\tau),\tau))d\tau
 \]
 \begin{equation}                                        \label{3.5}
 +\varphi_{i}(z_0^{}),\ \
 i=1,...,n, \ 0 \leq z_0 \leq L_0, \ t>0.
 \end{equation}
The following integral equation for $c(z_0^{},t)$ is derived from
(\ref{3.1}) and (\ref{2.3})
 \[
  c(z_0^{},t)= z_0^{} + \int_0^t d\tau
    \int_0^{z_0^{}}\sum_{i=1}^{n}((r_{M,i}^{}(c(\zeta_0^{},\tau),\tau,{\bf X}
    (c(\zeta_0^{},\tau),\tau),{\bf S}(c(\zeta_0^{},\tau),\tau))
 \]
 \begin{equation}                                       \label{3.6}
  + r_{i}^{}(c(\zeta_0^{},\tau),\tau,{\bf S}(c(\zeta_0^{},\tau),\tau),
  \mbox{\boldmath $\Psi$}(c(\zeta_0^{},\tau),\tau)))
    \frac{\partial c}{\partial \zeta_0}(\zeta_0,\tau)\ d\zeta_0, 0
    \leq z_0 \leq L_0,  t>0.
 \end{equation}
From (\ref{3.6}) it follows easily
 \[
 \frac{\partial c}{\partial z_0^{}}(z_0^{},t)\ = 1 +
 \int_0^{t}\sum_{i=1}^{n}((r_{M,i}^{}(c(z_0^{},\tau),\tau,{\bf X}(c(z_0^{},\tau),\tau),{\bf S}(c(z_0^{},\tau),\tau))
 \]
\begin{equation}                                       \label{3.7}
  + r_{i}^{}(c(z_0^{},\tau),\tau,{\bf S}(c(z_0^{},\tau),\tau),
  \mbox{\boldmath $\Psi$}(c(z_0^{},\tau),\tau)))
    \frac{\partial c}{\partial \zeta_0}(z_0,\tau) d\tau, 0 \leq z_0 \leq L_0, \ t>0.
\end{equation}
 The integral equations for $S_j(z,t)$ are obtained by integrating
 (\ref{2.11}) and considering the boundary conditions
 (\ref{2.9})

 \[
  S_j(z,t)= S_{j}^*(t) + D_{j}^{-1}\int_0^z(L-z)r_{S,j}^{}
  (\zeta,{\bf X}(\zeta,t),{\bf S}(\zeta,t))d\zeta
 \]
 \[
  +  D_{j}^{-1}\int_z^{L}(L-\zeta)r_{S,j}^{}(\zeta,{\bf X}(\zeta,t),{\bf S}
  (\zeta,t))d\zeta
 \]
 \begin{equation}                                       \label{3.8}
  +  \frac{h}{D_{j}^*}\int_0^{L}r_{S,j}^{}(\zeta,{\bf X}(\zeta,t),{\bf S}
  (\zeta,t))d\zeta, \ j=1,...,m, \ \ 0\leq z \leq L(t), \ t>0.
 \end{equation}
Similarly, the following integral equations for $\Psi_i$ are obtained
\[
  \Psi_i(z,t)= \psi_i^*(t) + D_{M,i}^{-1}\int_0^z(L-z)r_{i}^{}
  (\zeta,{\bf S}(\zeta,t),\mbox{\boldmath $\Psi$}(\zeta,t))d\zeta
\]
\begin{equation}                                       \label{3.9}
    + D_{M,i}^{-1}\int_z^{L}(L-\zeta)r_{i}^{}(\zeta,{\bf S}(\zeta,t),\mbox{\boldmath $\Psi$}(\zeta,t))d\zeta
  \ \ \ i=1,...,n, \ \ 0 \leq z\leq L(t), \ t>0.
\end{equation}
From (\ref{3.8}) it follows
 \begin{equation}                                       \label{3.10}
     \frac{\partial S_j}{\partial z} (L,t)=-D_{j}^{-1}
     \int_0^{L}r_{S,j}(\zeta,{\bf X}(\zeta,t),{\bf S}(\zeta,t))d\
     \zeta.
 \end{equation}
 Considering (\ref{3.10}) in (\ref{2.10}), equation for $S_j^*(t)$ writes
 \[
     \dot S_j^*(t)=(A/V)
     \int_0^{L}r_{S,j}(\zeta,t),{\bf X}(\zeta,t),{\bf S}(\zeta,t))d\zeta +
     (Q/V)(S_j^{in}-S_j^*(t)).
 \]
Integrating the last equation over time leads to the following integral
equation for $S_j^*(t)$
\[
S_j^*(t)=  \int_0^t\exp(-Q(t-\tau)/V)d\tau
\int_0^{L}(A/V)r_{S,j}^{}(\zeta,{\bf X}(\zeta,\tau),{\bf S}(\zeta,\tau))
     d\zeta
\]
\begin{equation}                                        \label{3.11}
+S_j^{in}, \ \ j=1,...,m,\ t>0.
\end{equation}
Following the same reasoning, a similar equation is obtained for $\psi_i^*(t)$
\[
 \psi_i^*(t)=  \int_0^t\exp(-Q(t-\tau)/V)d\tau
\int_0^{L}(A/V)r_{\psi,i}^{} (\zeta,{\bf S}(\zeta,\tau),
 \mbox{\boldmath $\Psi$}(\zeta,\tau))     d\zeta
\]
\begin{equation}                                        \label{3.12}
        +\psi_i^{in}, \ \ i=1,...,n,\ t>0.
\end{equation}
The integral equation for $L(t)$ is obtained from (\ref{2.4})
\begin{equation}                                        \label{3.13}
   L(t)= L_0+\int_0^t u(L(\tau),\tau)\ d\tau
     -\int_0^t\sigma_d(L(\tau))\ d\tau, \ t>0.
 \end{equation}

 Let us note that, as outlined at the end of Section \ref{n2}, the
 integral equations above depend on time and the space variable
 $z_0$ defined in the fixed domain $0\leq z\leq  L_0$. This result is
 essential to prove the existence and uniqueness of solutions. Indeed, following
 \cite{reactor,inv2}, a suitable contractive map can be introduced in the
 space of continuous functions and the fixed point theorem
 can be applied. We neglect the calculations since they are a generalization
 of \cite{reactor,inv2} with small modifications.


 \section{Anammox invasion model}\label{n4}
 \setcounter{equation}{0} \setcounter{figure}{0}

  In the previous sections, we performed the qualitative analysis
  for the invasion free boundary value problem of a biofilm reactor model.
  In particular, a result on the existence and uniqueness of solutions
  was provided. However, it is apparent that when complex biological cases are discussed,
  only numerical simulations can provide satisfactory predictions. The previous qualitative analysis gives a
  solid base to calculations. For the numerical solution of the model we use an extension of the numerical 
	method proposed in \cite{dac1}. The code is implemented in MatLab platform and simulations are run for a 
	set target simulation time $T$ that will be specified later on.
	
	The simulated biofilm system consists of bacterial cells accumulating on
  a surface surrounded by an aquatic region and reproduces a typical
  multi-culture and multi-substrate process which
  establishes in the deammonification units of the wastewater treatment
  plants. The deammonification process consists in the autotrophic nitrogen removal
  carried out by two microbial groups, the ammonium oxidizing bacteria
  AOB $(X_1)$ which oxidize ammonium $S_1$ partially to nitrite $S_2$
  aerobically and the anaerobic
  ammonium oxidizing bacteria AMX $(X_2)$, which subsequently convert
  the remaining ammonium and the newly formed nitrite into nitrogen gas
  and nitrate $S_3$ in trace concentrations. This process is also known as partial
  nitritation/anammox \cite{cao}.
  In multispecies biofilms, the AOB and AMX compete with other two
  major microbial groups: the nitrite oxidizing bacteria
  NOB $(X_3)$, which oxidize $S_2$ to $S_3$ in aerobic conditions and
  compete with $X_1$ for oxygen and $X_2$ for nitrite,
  and heterotrophic bacteria HB $(X_4)$. The latter can be further classified
  in ordinary heterotrophic organisms oxidizing the organic material and
  denitrifiers reducing nitrate to nitrite and nitrite to dinitrogen gas
  by consuming organic substrate $S_4$. $X_4$ compete with $X_1$ and $X_3$
  for oxygen $S_5$ and with $X_2$ for nitrite, the limiting substrate of
  $X_2$ in most instances.
  The establishment and proliferation of $X_2$ in such constituted biofilms strictly depends
  on the formation of an anoxic zone in the inner parts of the matrix where
  $X_3$ cannot grow, due to oxygen limitation.
	
	The mathematical model takes into consideration the dynamics of the five
  microbial species $X_i(z,t)$, including inert material $X_5$ which derives
  from microbial decay,
  and the five reactive components $S_j(z,t)$ within the biofilm. The
  corresponding concentrations in the bulk liquid $S_j^*(t)$ are
  taken into account as well.
  Planktonic cells have been considered for both $X_2$ and $X_4$ species as
  the model is aimed at simulating the invasion of a constituted biofilm
  by heterotrophic and Anammox bacteria
  after the establishment of a favorable environmental niche. Two
  modelling scenarios have been considered: the case of $X_2$ as
  single invading species and the case of $X_2$ and $X_4$ invasion and establishment
  in an autotrophic biofilm. Hereafter, they will be referred as
  Model 1, considered in  Section \ref{n4.1}, and Model 2 that will be discussed
  in Section \ref{n4.2}.
	
	\subsection{Model 1 -- One invading species}\label{n4.1}

\begin{figure*}
 \includegraphics[width=0.6\textwidth]{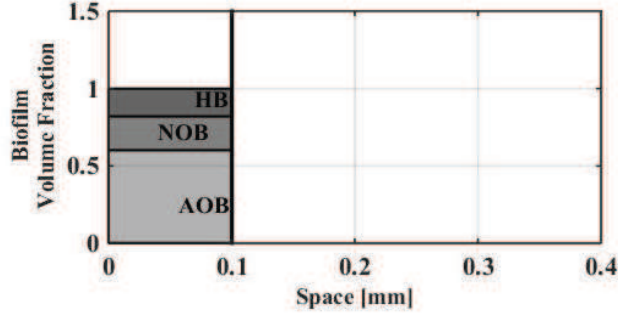}
 \caption{Initial biofilm configuration for Model 1.}
 \label{f4.1}
 \end{figure*}

  Model 1 considers a single invading species: the anaerobic
  ammonium oxidizing bacteria AMX $(X_2)$. The mathematical formalization of the problem is provided below. 
	The microbial species dynamics is governed by  equations (\ref{2.1}) rewritten in terms of $f_i$ for convenience

  \begin{equation}                                        \label{4.1}
     \frac{\partial f_i}{\partial t} +\frac{\partial}{\partial z}(u f_i)
     = r_{M,i}^{}(z,t,{\bf X},{\bf S}) + r_{i}^{}(z,t,{\bf S},
        \mbox{\boldmath $\Psi$}), \ i=1,...,5.
 \end{equation}
 The following initial volume fractions are associated to
 equations (\ref{4.1})
\begin{equation}                                                         \label{4.2}
   f_1(z,0)=0.65,\ f_2(z,0)=0.0, \ f_3(z,0)=0.25,\ f_4(z,0)=0.1,\
 f_5(z,0)=0.0.
\end{equation}
 The biofilm is assumed to be initially
 constituted only by the species $X_1$, $X_3$, $X_4$.
 The invasion of the species $X_2$ is simulated.
 The initial biofilm thickness $L_0$ is given by
 \begin{equation}                                                         \label{4.3}
   L_0= 0.1\ mm.
\end{equation}
A representation of the initial microbial distribution is reported
in Fig. \ref{f4.1}.

The net specific biomass growth rates $r_{M,i}, \ i=1,...,4$ are expressed as

 \[
 r_{M,i}^{}= (\mu_i^{}(\textbf{S})-k_{d,i})f_i,
\]

\noindent
where the function $\mu_i^{}(\textbf{S})$ denotes the $ith$ biomass specific growth rate and depends on the anabolic reactions performed by the $ith$ microbial species. 
It is usually formulated as Monod kinetics as detailed below. The term $k_{d,i}$ accounts instead for the forms of biomass loss and
energy requirements not associated with growth, including decay, maintenance, endogenous respiration, lysis, predation,
death. The net specific growth rates associated to $X_i, \ i=1,...,4$ are the the following

\begin{equation}                                        \label{4.4}
 r_{M,1}^{}= (\mu_1^{}(\textbf{S})-k_{d,1})f_1 =\left(\mu_{\max,1}\frac{S_1}{K_{1,1}+S_1}\frac{S_5}{K_{1,5}+S_5} - k_{d,1}\right)f_1,
 \end{equation}

 \begin{equation}                                        \label{4.5}
 r_{M,2}^{}= (\mu_2^{}(\textbf{S})-k_{d,2})f_2 = \left(\mu_{\max,2}\frac{K_{2,5}}{K_{2,5}+S_5}\frac{S_1}{K_{2,1}+S_1}\frac{S_2}{K_{2,2}+S_2}-k_{d,2}\right)f_2,
 \end{equation}

 \begin{equation}                                        \label{4.6}
 r_{M,3}^{}= (\mu_3^{}(\textbf{S})-k_{d,3})f_3 = \left(\mu_{\max,3}\frac{S_2}{K_{3,2}+S_2}\frac{S_5}{K_{3,5}+S_5}-k_{d,3}\right)f_3,
 \end{equation}

 \[
 r_{M,4}^{}= (\mu_{4,1}^{}(\textbf{S})+\mu_{4,2}^{}(\textbf{S})+\mu_{4,3}^{}(\textbf{S})-k_{d,4})f_4
 \]
 \[
 = \left(\mu_{\max,4}\frac{S_4}{K_{4,4}+S_4}\frac{S_5}{K_{4,5}+S_5} + \beta_1 \mu_{\max,4}\frac{K_{4,5}}{K_{4,5}+S_5}
             \frac{S_4}{K_{4,4}+S_4}\frac{S_3}{K_{4,3}+S_3}\frac{S_3}{S_{2}+S_3}\right. 
 \]						
 \begin{equation}                                        \label{4.7}						
						\left. + \beta_2 \mu_{\max,4}\frac{K_{4,5}}{K_{4,5}+S_5}\frac{S_4}{K_{4,4}+S_4}\frac{S_2}{K_{4,2}+S_2}\frac{S_2}{S_{3}+S_2}-k_{d,4}\right)f_4,
 \end{equation}

 \noindent
 where $\mu_{\max,i}$ denotes the maximum net growth rate for biomass $i$, $K_{i,j}$ the affinity constant of substrate $j$ for biomass $i$, $\beta_1$ and $\beta_2$
 the reduction factor for denitrification nitrate to nitrite and nitrite to nitrogen gas respectively.

 The autotrophic performance in the deammonification process relies on the activity of $X_1$ and $X_2$ and results in the $S_1$ conversion to dinitrogen gas via $S_2$.
 In aerobic environments, $S_1$ represents the preferential substrate for $X_1$ growth (\ref{4.4}). $X_2$ proliferate, in turn, on $S_1$ and $S_2$ and their metabolic activity
 is strongly affected by the oxygen concentration, the latter being inhibitory even at low concentrations (\ref{4.5}). Moreover, they rely on the production of $S_2$ by $X_1$, when that substrate
 is not provided from the bulk liquid. $X_3$ oxidize $S_2$ to $S_3$ under aerobic conditions and thus they compete with $X_2$ for $S_2$ (\ref{4.6}).
 $X_4$ are considered facultative bacteria: they can aerobically oxidize the organic matter ($\mu_{4,1}$) or perform
 denitrification reactions over $S_3$ and $S_2$ ($\mu_{4,2}$ and $\mu_{4,3}$ respectively). Indeed, in presence of $S_4$, $S_2$ and $S_3$ can be contextually consumed by $X_4$ according to
 equation (\ref{4.7}). In particular, $S_3$ and $S_2$ are reduced to dinitrogen gas in a sequential process which first converts $S_3$ into $S_2$, the latter being then reduced to $N_2$.
 In addition, the ratios $S_3/(S_{2}+S_{3})$ and $S_2/(S_{2}+S_{3})$ varying between $0\div 1$, indicate the percentage of biomass growing on nitrate and/or nitrite respectively.
 Inert has been treated as an additional microbial species whose growth rate depends on the decay of all the active species

 \begin{equation}                                        \label{4.8}
 r_{M,5}^{}=k_{d,1}f_1+k_{d,2}f_2+k_{d,3}f_3+k_{d,4}f_4.
 \end{equation}

 The specific growth rates $r_i^{}$ induced by the switch of the planktonic cells to
 the sessile mode of growth are defined as

 \begin{equation}                                        \label{4.9}
 r_{1}^{}=r_{3}^{}=r_{4}^{}=r_{5}^{}=0,
 \end{equation}

 \begin{equation}                                        \label{4.10}
 r_{2}^{}=k_{col,2}\frac{\Psi_2}{k_{\psi,2}+\Psi_2}
 \frac{K_{2,5}}{K_{2,5}+S_5}\frac{S_1}{K_{2,1}+S_1}\frac{S_2}{K_{2,2}+S_2}.
 \end{equation}

 Note that the growth rate terms $r_2^{}$ for $X_2$ is newly introduced as Monod
kinetics and indicate that the transition of bacteria from planktonic
state $\psi_2$ into the sessile state $X_2$
is controlled by the formation of a specific environmental niche which is
strictly connected to the local concentration of dissolved substrates.
The presence of planktonic species
is fundamental for the occurrence of the invasion process, as better specified
 in the following
 \vspace{1mm}

 {\bf Remark 1.} Consider the second equation in (\ref{2.8}) with $r_{M,2}$
 given by  (\ref{4.5}) and initial condition $f_2(z,0)=0$. If it is supposed that
 $r_{2}^{}=0$, then the mentioned equation admits the unique solution
 $f_2(z,t)=0$ and the species  $X_2$ cannot develop.
 \vspace{1mm}

 The diffusion of substrates is governed by
 \begin{equation}                                        \label{4.11}
   \frac{\partial S_j}{\partial t}-D_j\frac{\partial^2 S_j}{\partial z^2}=
   r_{S,j}^{}(z,t,{\bf X},{\bf S}),\
   j=1,...,5,
 \end{equation}
 with  the following initial-boundary conditions
\begin{equation}                                                         \label{4.12}
   S_j(z,0)=0,\ \frac{\partial S_j}{\partial z}(0,t)=0,\ j=1,...,5,
\end{equation}
\begin{equation}                                                         \label{4.13}
 S_j(L(t),t)=S_{j}^*(t),  \ j=1,...,4,
 \ S_5(L(t),t)=\overline{S}_5=1.5\ mg O_2/L.
\end{equation}
 The last condition simulates a continuous aeration of the biofilm reactor, \cite{masic}.

 The net substrate conversion rates account for both the microbial production and consumption (positive and negative terms respectively) and can
 be formulated from the corresponding microbial growth rates through the specific microbial yield $Y_i$. 
 They are usually expressed as double-Monod kinetics as presented below.

The ammonium conversion rate $r_{S,1}^{}$ is expressed as
\begin{equation}                                        \label{4.14}
r_{S,1}^{}=(-\frac{1}{Y_1}-i_{N,B}^{})\mu_1^{}X_1+
           (-\frac{1}{Y_2}-i_{N,B}^{})\mu_2^{}X_2
           -i_{N,B}^{}(\mu_{3}^{}X_3+\mu_{4,1}^{}X_4+\mu_{4,2}^{}X_4+\mu_{4,3}^{}X_4),
\end{equation}

\noindent
where $Y_i$ denotes the yield for biomass $i$ and $i_{N,B}^{}$ is the nitrogen content in biomass.
Ammonium can be directly consumed by AOB and AMX (first and second term in \ref{4.14}), and it is usually uptaken by other microbial species for anabolic reactions 
(third term in \ref{4.14}).

The nitrite and nitrate conversion rates $r_{S,2}^{}$ and $r_{S,3}^{}$ can be written as 

\begin{equation}                                        \label{4.15}
r_{S,2}^{}=  \frac{1}{Y_1}\mu_1^{}X_1
             -(\frac{1}{Y_2}+\frac{1}{1.14})\mu_2^{}X_2
             -\frac{1}{Y_3}\mu_{3}^{}X_3
            -(1-\frac{1}{Y_4})\frac{1}{1.14}\mu_{4,2}^{}X_4
						+(1-\frac{1}{Y_4})\frac{1}{1.72}\mu_{4,3}^{}X_4,
\end{equation}

\begin{equation}                                        \label{4.16}
r_{S,3}^{}= (\frac{1}{1.14})\mu_2^{}X_2+
             +\frac{1}{Y_3}\mu_{3}^{}X_3
             + (1-\frac{1}{Y_4})\frac{1}{1.14}\mu_{4,2}^{}X_4.
\end{equation}

In aerobic environments, nitrite $S_2$ are produced through the ammonium conversion catalyzed by AOB and they are further oxidized to nitrate $S_3$ by NOB (first and third term in \ref{4.15}).
Obviously, the latter represents a production rate for nitrate (second term in \ref{4.16})
Conversely, in anoxic conditions AMX bacteria convert nitrite and ammonium in dinitrogen gas (second term in \ref{4.15}), 
while HB consume both the oxidized nitrogen compounds by reducing nitrate to nitrite (fourth and third term in \ref{4.15} and \ref{4.16} respectively) 
and by using nitrite as oxygen source (fifth term in \ref{4.15}).

The organic carbon conversion rate $r_{S,4}$ is expressed by

\begin{equation}                                        \label{4.17}
r_{S,4}^{}=
-\frac{1}{Y_4}(\mu_{4,1}^{}X_4+\mu_{4,2}^{}X_4+\mu_{4,3}^{}X_4)
\end{equation}

\noindent
and indicates the $S_4$ consumption due to $X_4$ metabolism in both aerobic and anoxic conditions.

Finally, $r_{S,5}$ describes the oxygen conversion rate within the multispecies biofilm

\begin{equation}                                        \label{4.18}
r_{S,5}^{}=(1-\frac{3.43}{Y_1})\mu_1^{}X_1
            +(1-\frac{1.14}{Y_3})\mu_3^{}X_3
            +(1-\frac{1}{Y_4})\mu_{4,1}^{}X_4,
\end{equation}

where the three terms in (\ref{4.18}) are net consumption rates due to AOB, NOB and HB species using oxygen for their metabolisms.

The functions $S_j^*(t)$ are governed by the following initial
 value problem for ordinary differential equations
 \begin{equation}                                        \label{4.19}
     V\dot S_j^*=-AD_{j} \frac{\partial S_j}{\partial z}(L(t),t)+
   Q(S_j^{in}-S_j^*(t)),\    j=1,...,4,
 \end{equation}
 The initial conditions for $S_j^*$ are the following
 \begin{equation}                                                  \label{4.20}
 S_1^{in}=1200\ mg N/L,\ S_2^{in}=S_3^{in}=0,\ S_4^{in}=120\ mg
 COD/L.
\end{equation}
 The inlet concentrations are non-zero only for $S_1$ and $S_4$,
 reproducing the case of a biofilm reactor fed with a wastewater
 containing both ammonium nitrogen and organic carbon.

The diffusion and reaction of planktonic cells within the biofilm
 matrix is governed by the following equations
 \begin{equation}                                        \label{4.21}
   \frac{\partial \Psi_i}{\partial t}-D_{M,i}\frac{\partial^2 \Psi_i}{\partial z^2}=
   r_{\Psi,i}^{}(z,t,{\bf S},\mbox{\boldmath $\Psi$}),\
   i=1,...,5,
 \end{equation}
 where $D_{M,i}$ denotes the diffusivity coefficient.
 The conversion rates of planktonic cells  due
 to invasion process
 are expressed by
\begin{equation}                                        \label{4.22}
r_{\psi,i}^{}=-\frac{1}{Y_{\psi,i}}r_i^{},\ i=1,...,5,
\end{equation}
with $Y_{\psi,i}$ being the yield of sessile species on planktonic ones.
 They are assumed proportional to $r_i^{}$, i.e. described
by the same Monod kinetics \cite{inv}. The following initial-boundary conditions
 are associated to equations (\ref{4.21})
  \begin{equation}                                          \label{4.23}
 \Psi_i(z,0)=0,\ \frac{\partial \Psi_i}{\partial z}(0,t)=0,
 \Psi_i(L(t),t)=\psi_{i}^*(t),\ i=1,...,5.
\end{equation}
 The functions $\psi_i^*(t)$ satisfy
  the following initial
 value problem for ordinary differential equations
 \begin{equation}                                        \label{4.24}
     V\dot \psi_i^*=-AD_{M,i}^{}
   \frac{\partial \Psi_i}{\partial z}(L(t),t)+
   Q(\psi_i^{in}-\psi_i^*(t)),\
    i=1,...,5,
 \end{equation}
 \begin{equation}                                        \label{4.25}
    \psi_1^{in}=0,\   \psi_2^{in}=1.0\ mg COD/L,\
        \psi_3^{in}=\psi_4^{in}=\psi_5^{in}=0.
 \end{equation}

 Note that for $i=1$ equation for $\Psi_1$ is homogeneous because of hypothesis
 (\ref{4.9}) and equation for $\psi_1^*$ do not contain the term
 $\psi_1^{in}$ because of hypothesis
 (\ref{4.25}). Therefore, the system of the two equations admits the unique solution
 $\Psi_1(z,t)=0$, $\psi_1^*(t)=0$. Same result holds for
 $\Psi_3(z,t)=0$, $\psi_3^*(t)=0$, $\Psi_4(z,t)=0$, $\psi_4^*(t)=0$,
 $\Psi_5(z,t)=0$, $\psi_5^*(t)=0$.

 The biofilm reactor is
 characterized by the following operational parameters: the flow
 rate $Q$ is set to $3.15 L/d$, the surface area available for
 biofilm attachment and proliferation $A$ is equal to $1 m^2$ and
 the reactor volume is of $3.15 L$, leading to a hydraulic
 retention time of 1 day.

 The values of the stoichiometric and kinetic parameters used for numerical
 simulations are adopted from \cite{mattei2} and are reported for convenience
 in Tables \ref{T1}-\ref{T1bis}.

 \begin{table}
\begin{footnotesize}
 \begin{center}
 \begin{tabular}{lccccc}
 \hline
{\textbf{Symbol}} & {\textbf{Definition}} &  \textbf{Value} &  \textbf{Units} \\
 \hline
 $Y_1$           & $X_1$ yield on $S_1$                       & 0.150 & $g COD/g N$       \\
 $Y_2$           & $X_2$ yield on $S_1$                       & 0.159 & $g COD/g N$       \\
 $Y_3$           & $X_3$ yield on $S_1$                       & 0.041 & $g COD/g N$       \\
 $Y_4$           & $X_4$ yield on $S_4$                       & 0.63  & $g COD/g COD$     \\
 $\mu_{\max,1}$  & Maximum growth rate of $X_1$               & 2.05  & $d^{-1}$          \\
 $\mu_{\max,2}$  & Maximum growth rate of $X_2$               & 0.08  & $d^{-1}$          \\
 $\mu_{\max,3}$  & Maximum growth rate of $X_3$               & 1.45  & $d^{-1}$          \\
 $\mu_{\max,4}$& Maximum growth rate of $X_4$                 & 6.0   & $d^{-1}$          \\
 $K_{1,1}$       & $S_1$ affinity constant for $X_1$          & 2.4   & $mg N/L$      \\
 $K_{1,5}$       & $S_5$ affinity constant for $X_1$          & 0.6   & $mg O_2/L$    \\
 $K_{2,1}$       & $S_1$ affinity constant for $X_2$          & 0.07  & $mg N/L$      \\
 $K_{2,2}$       & $S_2$ affinity constant for $X_2$          & 0.05  & $mg N/L$      \\
 $K_{2,5}$       & $S_5$ inhibiting constant for $X_2$        & 0.01  & $mg O_2/L$    \\
 $K_{3,2}$       & $S_2$ affinity constant for $X_3$          & 5.5   & $mg N/L$      \\
 $K_{3,5}$       & $S_5$ affinity constant for $X_3$          & 2.2   & $mg O_2/L$    \\
 $K_{4,4}$       & $S_4$ affinity constant for $X_4$          & 4.0   & $mg COD/L$    \\
 $K_{4,5}$       & $S_5$ affinity/inhibiting constant for $X_4$  & 0.2   & $mg O_2/L$ \\
 $K_{4,2}$       & $S_2$ affinity constant for $X_4$          & 0.5   & $mg N/L$      \\
 $K_{4,3}$       & $S_3$ affinity constant for $X_4$          & 0.5  & $mg N/L$       \\
 $k_{d,1}$       & Decay constant for $X_1$                   & 0.0068  & $d^{-1}$    \\
 $k_{d,2}$       & Decay constant for $X_2$                   & 0.0026  & $d^{-1}$    \\
 $k_{d,3}$       & Decay constant for $X_3$                   & 0.04    & $d^{-1}$    \\
 $k_{d,4}$       & Decay constant for $X_4$                   & 0.06    & $d^{-1}$    \\
 \hline
 \end{tabular}
 \caption{Kinetic and Stoichiometric Parameters used for Numerical Simulations} \label{T1}
 \end{center}
 \end{footnotesize}
 \end{table}

 \begin{table}
\begin{footnotesize}
 \begin{center}
 \begin{tabular}{cccccc}

 \hline
{\textbf{Symbol}} & {\textbf{Definition}} &  \textbf{Value} &  \textbf{Units} \\
 \hline
$i_{N,B}^{}$    & N content of biomass                       & 0.07  & $g N/g COD$       \\
$\beta_1$        & Reduction factor for denitrification $NO_3^{}-NO_2^{}$         & 0.8   & -                         \\
$\beta_2$        & Reduction factor for denitrification $NO_2^{}-N_2^{}$          & 0.8   & -                          \\
$k_{col,2}$      & Maximum colonization rate of $\psi_2$                          & 0.0001   & $d^{-1}$                   \\
$k_{col,4}$      & Maximum colonization rate of $\psi_4$                          & 0.0001   & $d^{-1}$                   \\
$Y_{\psi,2}$     & Yield of $X_2$ on $\psi_2$                                     & 0.001   & -                   \\
$Y_{\psi,4}$     & Yield of $X_4$ on $\psi_4$                                     & 0.001   & -                  \\
$k_{\psi,2}$     & Kinetic constant for $\psi_2$                                  & 0.000001   & $mg COD/L$                   \\
$k_{\psi,4}$     & Kinetic constant for $\psi_4$                                  & 0.000001   & $mg COD/L$                   \\
\hline
 \end{tabular}
 \caption{Kinetic and Stoichiometric Parameters used for Numerical Simulations} \label{T1bis}
 \end{center}
 \end{footnotesize}
 \end{table}

\begin{figure*}
\includegraphics[width=1\textwidth]{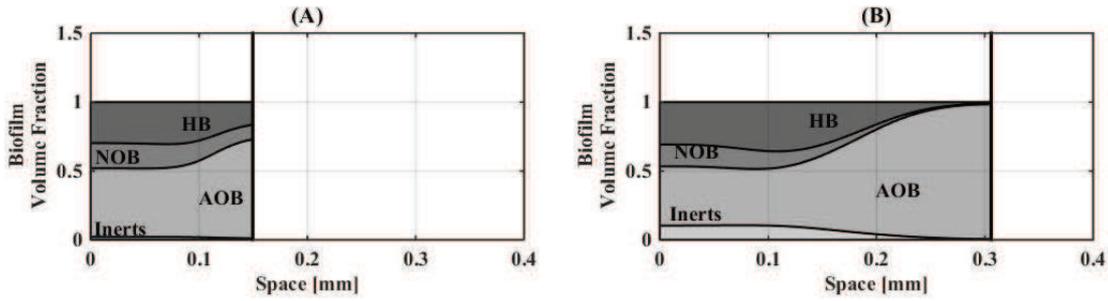}
 \caption{Microbial species distribution of a multispecies biofilm
 undergoing $\psi_2$ colonization after 1(A) and 5(B)
 days simulation time.} \label{f4.2}
 \end{figure*}

\begin{figure*}
\includegraphics[width=1\textwidth]{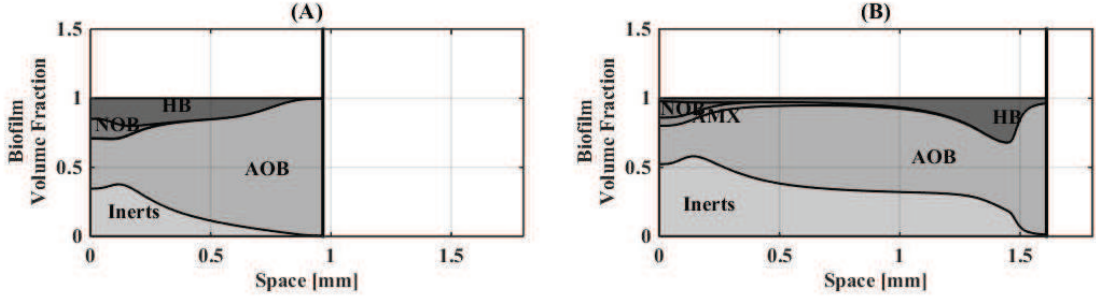}
 \caption{Microbial species distribution of a multispecies biofilm
 undergoing $\psi_2$ colonization after 20(A) and 50(B)
 days simulation time.} \label{f4.3}
 \end{figure*}

\begin{figure*}
\includegraphics[width=1\textwidth]{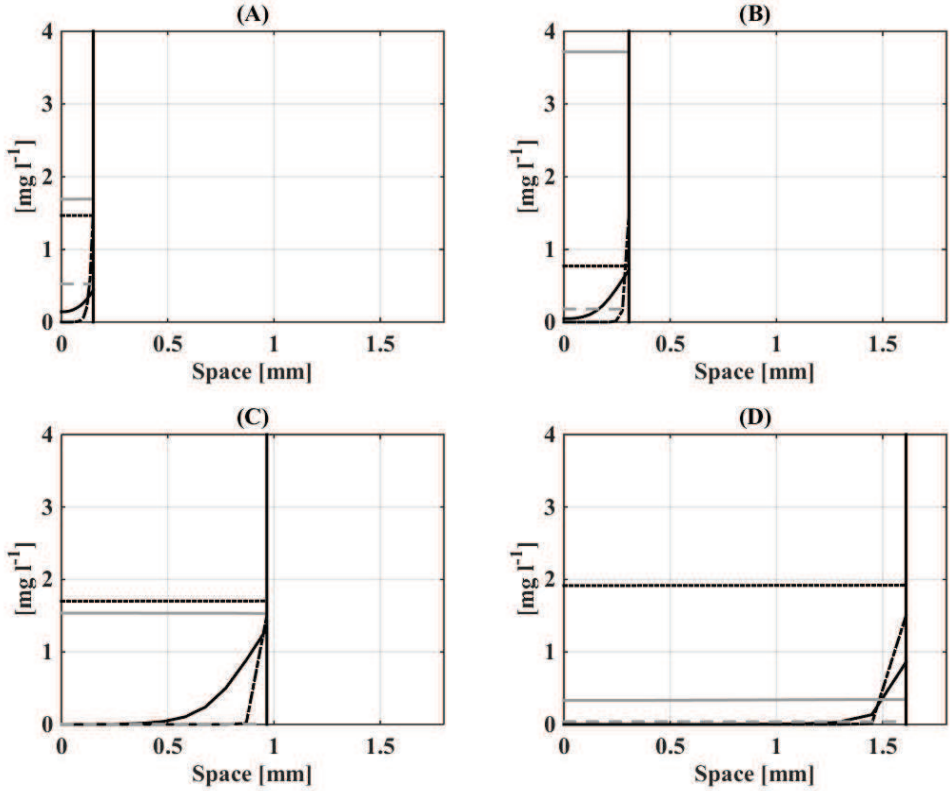}
 \caption{Substrate trends within a
 multispecies biofilm undergoing $\psi_2$ colonization after 1(A),
 5(B), 20(C) and 50(D) days simulation time. Dotted black line $S_1$, continuous grey line $S_2$,
 dashed grey line $S_3$, continuous black line $S_4$, dashed-dotted black line $S_5$. $S_1,S_2,S_3$ 
 concentrations are reduced by a factor of 0.002, 0.005 and 0.005 respectively.} \label{f4.4}
 \end{figure*}

The simulation results for the multispecies biofilm performance when the AMX invasion is considered 
are reported in Figs. \ref{f4.2}-\ref{f4.4}.
After one day of simulation time (Fig. \ref{f4.2}-A), it is possible
to notice that the microbial distribution into the biofilm is still affected
by the initial conditions and the colonization phenomenon has not occurred yet.
After 5 days of simulation time (Fig. \ref{f4.2}-B), the biofilm experiences oxygen limitation, 
due to the low concentration maintained within the bulk liquid. As a consequence, the NOB concentration significantly 
decreases with respect to the initial fraction, with the AOB and HB being the two species proliferating the most.
The AOB activity is confirmed by the decrease in $S_1$ concentration within the bulk liquid with respect to the  
inlet concentration and a concurrent increase in $S_2$ concentration (Fig. \ref{f4.4}-A,B). Note that 
the latter keeps higher than $S_3$ concentration as the metabolic activity of NOB is limited by the low oxygen concentration.
The organic carbon is completely depleted within the biofilm and its concentration keeps lower than $1 mgl^{-1}$.

Figure \ref{f4.3}-A displays biofilm configuration after 20 days of system
operation: nevertheless the concomitant formation of an anoxic zone in the inner part of
the biofilm (Fig. \ref{f4.4}-C) and a non-zero $\psi_2$ concentration all
over the biofilm (data not shown), AMX have not yet established in sessile form. This might be due to the very slow growth rate of $X_2$. The biofilm is
dominated by $X_1$ while $X_5$ predominate in the inner layer. 
Substrate trends assume the following configuration:
$S_1$ increases in the bulk liquid due to the lower AOB activity in the outer art of the biofilm where $S_5$ is totally consumed 
and its depletion determines the formation of an anoxic zone. Moreover, $S_2$ represents the main abundant
product while $S_3$ and $S_4$ keep close to zero all over the biofilm (Fig. \ref{f4.4}-C).
At day 50 (Figs. \ref{f4.3}-B and \ref{f4.4}-D), AMX have colonized the
environmental niche which formed at the bottom of the biofilm
(Fig. \ref{f4.3}-B). AOB still dominate the aerobic zone
while NOB are confined to the internal layers. $S_3$ concentration is close to zero as the metabolism of NOB significantly slows down. 
Note that the availability of $S_5$ within the biofilm is strictly connected to the penetration depth whose decrease leads to 
an increasing anoxic zone (Fig. \ref{f4.4}-D). Furthermore, AMX grow only where favorable environmental conditions establish
despite the biofilm results fully penetrated by the
same bacteria in motile/colonizing form $\psi_2$ for all simulation
times (data not shown).

AMX invasion is significantly influenced by many parameters such as environmental factors (i.e. pH and temperature) and operational conditions (i.e. dilution rate, C/N ratio, 
aeration pattern). The main goals for the further computational studies are to determine how the invasion phenomenon is affected by the oxygen and organic carbon availability. 
For this reason we vary the concentration of oxygen in the bulk liquid $S_{5L}$ and the organic carbon concentration in the inlet $S_{4}^{in}$ in the range $[0.5 - 6]$ 
and $[120 - 750]$ respectively. In the following, we will refer to Secs. \ref{n4.1.1} and \ref{n4.1.2} for the applications with the variable oxygen and organic carbon. 
We assumed the initial condition reported in Fig. \ref{f4.1} for all the simulation studies. 

\subsubsection{Model 1 -- Assessment -A- Effect of oxygen concentration}\label{n4.1.1}

\begin{figure*}
\includegraphics[width=1\textwidth]{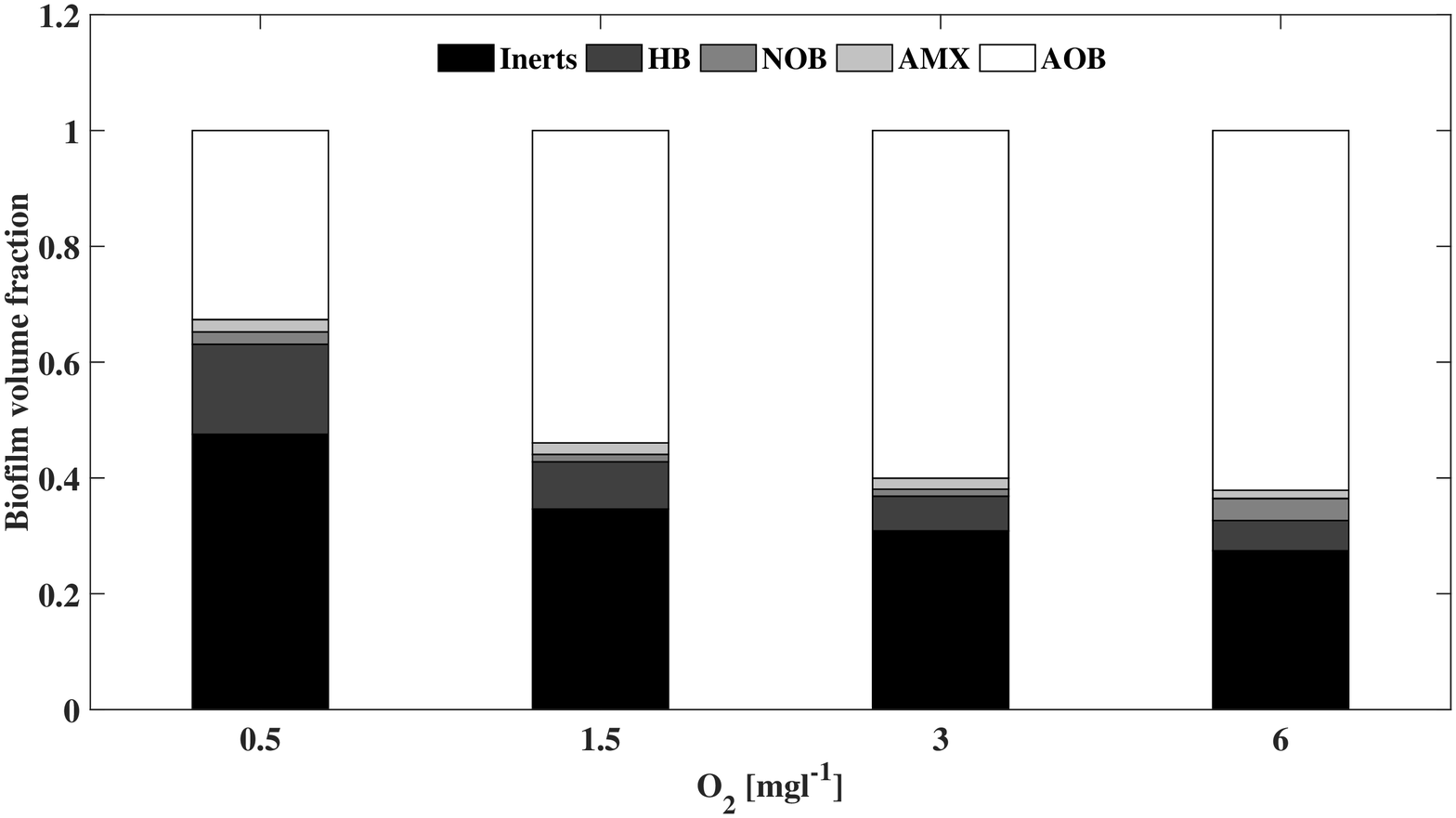}
 \caption{Total biofilm volume fractions at different $O_2$ concentrations after 50 days simulation time.} \label{f4.5}
 \end{figure*}

\begin{figure*}
\includegraphics[width=1\textwidth]{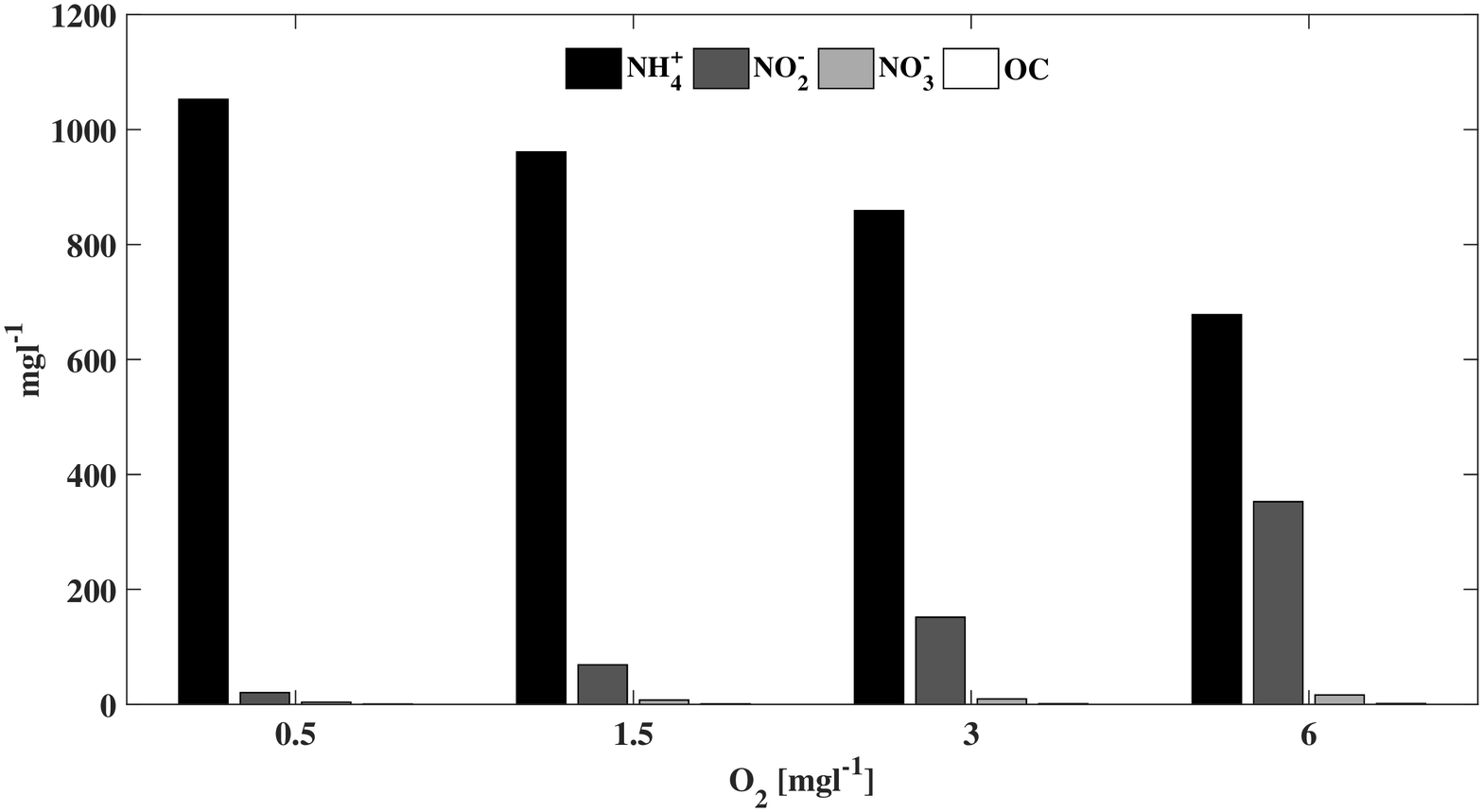}
 \caption{Substrate concentrations within the bulk liquid at different $O_2$ concentrations after 50 days simulation time.} \label{f4.6}
 \end{figure*}

Model outcomes for the simulation studies with variable oxygen are summarized in Figs. \ref{f4.5} and \ref{f4.6} 
in terms of biomass distribution and substrate concentrations within the bulk liquid. 
Four different oxygen levels (0.5 -- 1.5 -- 3 -- 6 $mgO_2/L$) have been tested and 
the simulations have been run for a target time of $50$ days.
AMX are strictly inhibited by the oxygen concentration and as expected, their total volume fraction
is found to slightly increase when varying the oxygen level from $6$ to $1.5$. A lower oxygen concentration 
leads to a decrease in AMX fraction; indeed, under this condition HB reaches the highest fraction 
competing with AMX and NOB for $S_2$. The optimal condition for AMX establishment and proliferation within the biofilm
occurs at $3$ $mgO_{2}/L$, even if the relative total biofilm fraction is lower with respect to $1.5$ $mgO_{2}/L$. 
Of course, NOB fraction is higher when the oxygen concentration is equal to $6$ $mgO_{2}/L$.
Regarding nitrogen removal, it is possible to note that the $S_1$ concentration 
progressively decreases and consequently $S_2$ increases going from $0.5$ to $6$ $mgO_2/L$ (Fig. \ref{4.6}).
These substrates show fully penetrated profiles (data not shown) and consequently the AMX can grow for all the cases with their maximum specific growth rate,
but only where anoxic conditions are established.

\subsubsection{Model 1 -- Assessment -B- Effect of inlet organic carbon concentration}\label{n4.1.2}

\begin{figure*}
\includegraphics[width=1\textwidth]{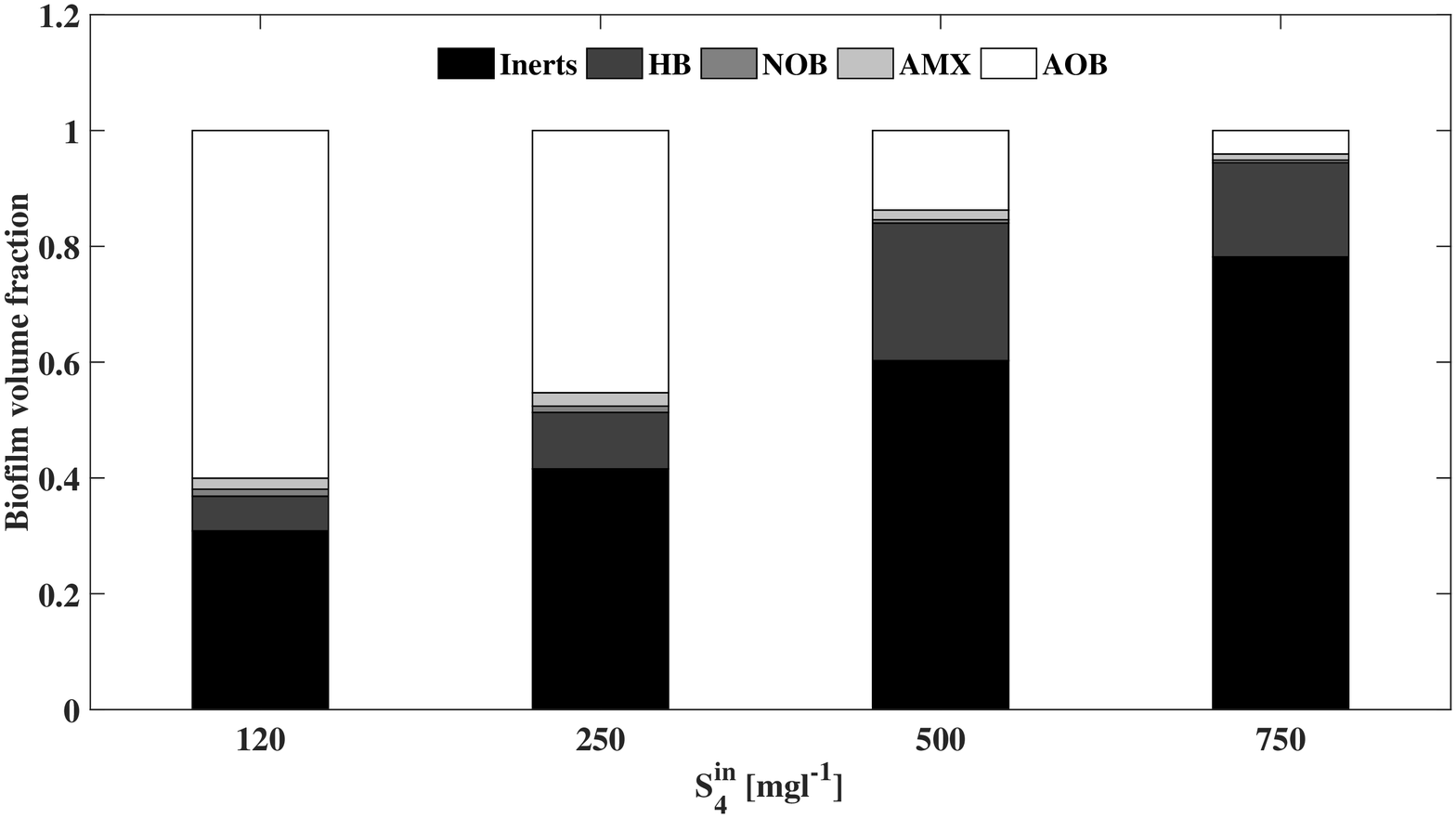}
 \caption{Total biofilm volume fractions at different $S_{4}^{in}$ values after 50 days simulation time.} \label{f4.7}
 \end{figure*}

\begin{figure*}
\includegraphics[width=1\textwidth]{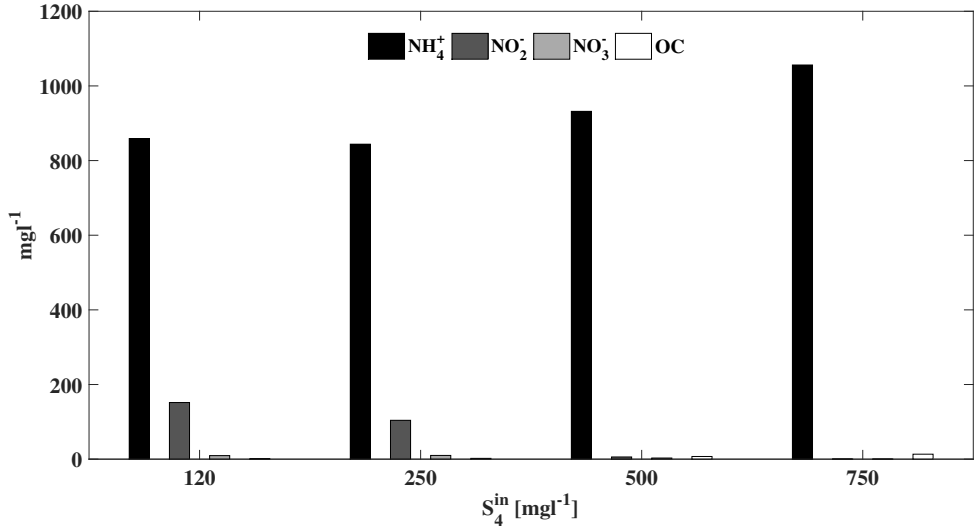}
 \caption{Substrate concentrations within the bulk liquid at different $S_{4}^{in}$ values after 50 days simulation time.} \label{f4.8}
 \end{figure*}

The second simulation studies investigated the effect of increasing C/N ratios on AMX performances. 
The oxygen concentration within the bulk liquid has been fixed to $3$ $mgO_{2}/L$.
As shown in Figs. \ref{f4.7} and \ref{f4.8}, four different concentrations 
of the inlet organic carbon $S_{4}^{in}$ have been tested (120 -- 250 -- 500 -- 750 $mgCOD/L$) and all the simulations have been run for $50$ days.
Figure \ref{f4.7} shows that the AOB volume fraction is prevalent when low organic carbon is available for HB, 
which compete for oxygen with all the other aerobic species in the external part of the biofilm. AMX invasion and proliferation 
is favored at $S_4^{in}=500 mgCOD/L$ since the NOB significantly decrease when increasing the inlet organic carbon concentration.
The highest carbon content leads to the highest inerts volume fraction as HB are strongly predominant and out-compete all the other species.

According to the volume fraction distribution, total nitrogen removal is higher when both AMX and AOB can easily perform their metabolisms while 
NOB activity is inhibited by HB. This particular condition is more evident when the inlet carbon concentration is $250$ $mgCODl^{-1}$ and a higher amount 
of dissolved oxygen is utilized by HB. Ammonium removal is not significant when both AMX and AOB are not prevalent within the biofilm and organic carbon removal starts 
to be incomplete when increasing $S_{4}^{in}$ concentration to $750$ $mgCODl^{-1}$.

\subsection{Model 2 -- Two invading species}\label{n4.2}

\begin{figure*}[ht]
\includegraphics[width=0.6\textwidth]{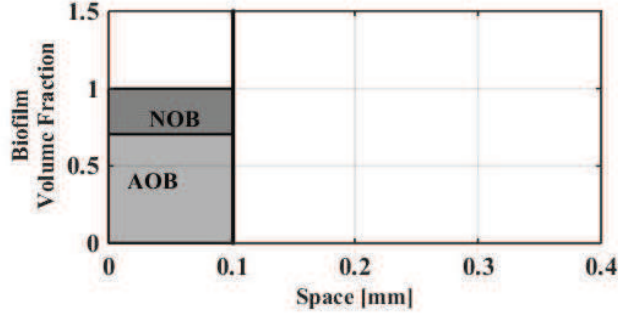}
 \caption{Initial biofilm configuration for Model 2.} \label{f4.9}
 \end{figure*}

In this section, the model was applied to the case of two species invasion, HB and AMX respectively.
The microbial species growth is governed by equations (\ref{4.1})
with the following initial volume fractions
\begin{equation}                                                  \label{4.26}
   f_1(z,0)=0.7,\ f_2(z,0)=0, \ f_3(z,0)=0.3,\ f_4(z,0)=0,\
 f_5(z,0)=0.
\end{equation}
 Only the species $X_1$ and $X_3$ are
 supposed to inhabit the biofilm at $t=0$.
 The invasion of the species $X_2$ and $X_4$ is simulated.
 The initial biofilm thickness $L_0$ is given by (\ref{4.3}).
 A representation of the initial microbial distribution is reported
 in Fig. \ref{f4.9}.

 The biomass growth rates $r_{M,i}$ are the same as Model 1, formulas
 (\ref{4.4})-(\ref{4.8}). The specific growth rates $r_{i}$ induced by the switch of
 the  planktonic cells to the sessile mode of growth are defined as

 \begin{equation}                                        \label{4.27}
 r_{1}^{}=r_{3}^{}=r_{5}^{}=0,
 \end{equation}
 \begin{equation}                                        \label{4.28}
 r_{2}^{}=k_{col,2}\frac{\Psi_2}{k_{\psi,2}+\Psi_2}
 \frac{K_{2,5}}{K_{2,5}+S_5}\frac{S_1}{K_{2,1}+S_1}\frac{S_2}{K_{2,2}+S_2},
 \end{equation}

 \[
 r_{4}^{}=k_{col,4}\frac{\Psi_4}{k_{\psi,4}+\Psi_4}\left(\frac{S_4}{K_{4,4}+S_4}\frac{S_5}{K_{4,5}+S_5} + \beta_1 \frac{K_{4,5}}{K_{4,5}+S_5}
\frac{S_4}{K_{4,4}+S_4}\frac{S_3}{K_{4,3}+S_3}\frac{S_3}{S_{2}+S_3} \right.
 \]
 \begin{equation}                                        \label{4.29}
             \left.
                        + \beta_2 \frac{K_{4,5}}{K_{4,5}+S_5}\frac{S_4}{K_{4,4}+S_4}\frac{S_2}{K_{4,2}+S_2}\frac{S_2}{S_{2}+S_3}\right).
 \end{equation}
 The growth rate terms $r_2^{}$ and  $r_4^{}$ for $X_2$ and
 $X_4$, respectively, indicate that the transition of
 bacteria from planktonic state $\psi_2^{}$,  $\psi_4^{}$ into
 the sessile state $X_2^{}$, $X_4^{}$
 is controlled by the formation of specific environmental niches
 connected to the local concentration of
 dissolved substrates. As in Model 1,
 consider second and fourth equation in (\ref{2.8}) with $r_{M,2}$
 and  $r_{M,4}$ given by (\ref{4.5}) and (\ref{4.7}), respectively.
 If it is supposed that
 $r_{2}^{}=r_{4}^{}=0$, then the mentioned equations with
 initial condition $f_2(z,0)=f_4(z,0)=0$
 admit the unique solution
 $f_2(z,t)=f_4(z,t)=0$ and the species  $X_2$ and $X_4$ cannot develop.

The initial-boundary conditions for $S_j$ and net conversion rates of substrates
 are the same as Model
 1, formulas (\ref{4.12})-(\ref{4.13}) and (\ref{4.14})-(\ref{4.18}),
  respectively. The
 initial conditions for $S_j^*$  are given by (\ref{4.20}).
 The initial-boundary conditions for $\Psi_i$ are same as Model 1,
 formula (\ref{4.23}).
 The initial conditions for $\psi_{i}^*$ are the following
 \begin{equation}                                             \label{4.30}
  \psi_1^{in}=0,\  \psi_2^{in}=1.0\ mg COD/L,\  \psi_3^{in}=0,\
    \psi_4^{in}=1.2\ mg COD/L,
  \  \psi_5^{in}=0.
 \end{equation}

Note that, by using the same arguments as Model 1, it can be shown that
 $\Psi_1(z,t)=0$, $\psi_1^*(t)=0$,
 $\Psi_3(z,t)=0$, $\psi_3^*(t)=0$,
 $\Psi_5(z,t)=0$, $\psi_5^*(t)=0$.

 The operational parameters of the biofilm reactor are the same as
 Model 1.
 
\begin{figure*}
 \includegraphics[width=1\textwidth]{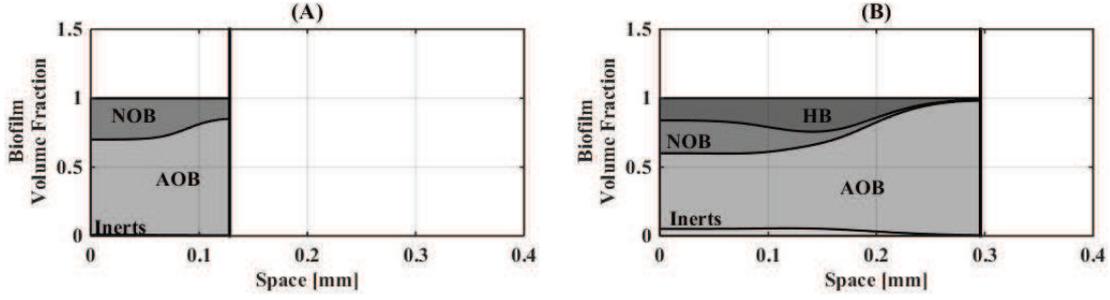}
  \caption{Microbial species distribution of a multispecies biofilm undergoing $\psi_2$ and $\psi_4$ colonization after 2(A) and 5(B) days simulation time.} \label{f4.10}
 \end{figure*}

\begin{figure}
\includegraphics[width=1\textwidth]{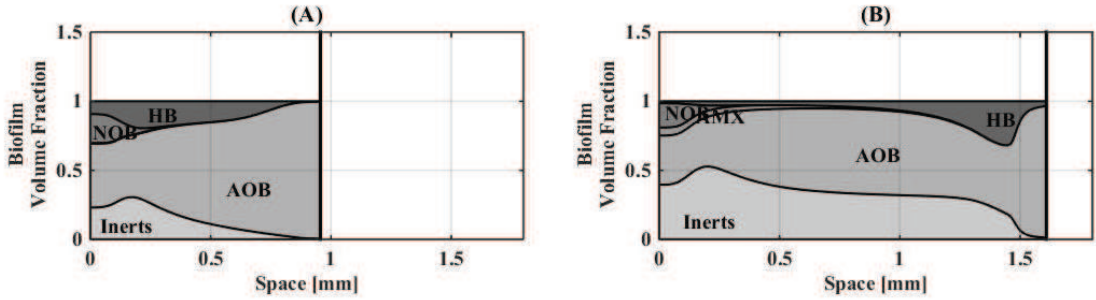}
  \caption{Microbial species distribution of a multispecies biofilm undergoing $\psi_2$ and $\psi_4$ colonization after 20(A) and 50(B) days simulation time.} \label{f4.11}
 \end{figure}

 In Figs. \ref{f4.10} and \ref{f4.11} the simulation results for the multispecies
 biofilm system with two invading species are reported. Differently from Model 1,
 the $X_4$ invasion is very fast and it is already visible after 5 days of simulation time
(Fig. \ref{f4.10}-B). This is due to the different environmental conditions
that trigger the invasion of the two microbial species.
Indeed the establishment of $X_2$ is dependent on the formation of an anoxic
zone within the biofilm while $X_4$ are facultative bacteria and can grow in
both aerobic and anoxic environments.
After 20 days the biofilm configuration is the same of the previous
application (Figs. \ref{f4.3}-A and \ref{f4.11}-A) and as we can expect,
the further evolution of the system is practically the same
for the two cases studied. Simulation results confirm model capability
of predicting the invasion phenomenon on time and space. Indeed, the
model is able to predict the delays between the $X_2$
and $X_4$ colonizations and the location where the two planktonic species
establish. To the best of our knowledge, such results cannot be achieved
by the existing continuum biofilm models but they might
have a significant impact on the developing of new strategies for
such biofilm reactors operation.

\section{Conclusion}\label{n6}

 In this work, the qualitative analysis of the free boundary problem
 related to the invasion phenomenon in biofilm reactors has been discussed.
 The model takes into account the dynamics of sessile species, nutrients and microbial products,
 and planktonic cells, the latter diffusing from the bulk liquid within the biofilm matrix, where they
 might switch their status from motile to sessile and thus colonize the pre-existing biofilm.
 The dynamics of bulk liquid have been explicitly modeled by considering two systems of
 nonlinear ordinary differential equations which derive from mass conservation principles. An existence and uniqueness result has been
 provided for the related free boundary value problem by using the method of characteristics and the fixed point theorem.
 It is important to notice that the planktonic species are just provided by the bulk liquid; however, the reverse process 
 which accounts for the switch from sessile to planktonic
 form of life might occur under specific conditions. This phenomenon could be explicitly taken into account by considering 
 a direct dependence of the free planktonic cell reaction rates on
 the concentration of the sessile bacteria. The same methodology adopted in this work could be easily adapted to address the existence and uniqueness questions of this new system.
 Numerical simulations related to a real biofilm system have been performed. Two specific model applications have been analyzed.
 Simulation results demonstrate the underlying conclusion that the invasion model can be effectively used as a predictive tool to develop
 specific reactor operation strategies. Further developments might be related to the definition of a calibration protocol through experimental data. 


\end{document}